\documentstyle[prb,aps,twocolumn]{revtex}
\begin{document}
\draft
\title{Comment on ``Local dimer-adatom stacking fault structures from 
$3\times3$ to $13\times13$ along Si(111)-$7\times7$ domain boundaries''}
\author{Makoto Itoh}
\address{Interdisciplinary Research Centre for Semiconductor Materials, \\
Imperial College, London SW7 2BZ, UK}
\maketitle
\begin{abstract}
Zhao {\it et al.} [\prb {\bf 58}, 13824 (1998)] depicted several atomic 
structures of domain boundaries on a Si(111) surface 
and criticized the article by the present author and the co-workers. 
I will point out that their criticism is incorrect and their structure models 
have no consistency. 
\end{abstract}

Zhao and co-workers referred to the work by the present author and the 
co-workers~\cite{itoh93} in the former part of their article~\cite{zhao98} and 
stated as if we merely tried to make a scheme to classify the domain 
boundaries (DBs) appearing on Si(111) surfaces with the $7\times7$ 
dimer-adatom-stacking-fault (DAS) reconstruction~\cite{taka85}. 
However, what we actually showed in Ref.~\onlinecite{itoh93} was that
any structures of a reconstructed Si(111) surface including those in the
DBs could be consistently accounted for by the atomic bondings 
which are already contained in the DAS structure. 
And this is why it was necessary to classify all the possible patterns 
of the DBs, followed by the structural analyses of the 14 patterns among 
the 27 possible ones. 
Then, we assumed the dimer-adatom interaction~\cite{payne,stich} 
and the reduction in the number of dangling bonds as the guiding principles 
to show that the atomic structures can be consistently determined 
with the use of the constituents of the DAS structure only.

In Ref.~\onlinecite{zhao98}, Zhao {\it et al.} criticized Fig.\ 6(b) of 
Ref.~\onlinecite{itoh93} by 
stating that we had given too many dangling bonds and, accordingly, they drew 
Fig.\ 1 in Ref.~\onlinecite{zhao98} by adding one dimer and one rest
atom in the DB in Fig.\ 6(b) of Ref.~\onlinecite{itoh93}. 
Yet, they did not add such dimers to the DB in Fig.\ 2, 
which is the structure model of their own scanning tunneling microscopy (STM) 
image. 
In contrast to one of our assumptions, the dimer they added in Fig.\ 1 is not 
accompanied by any adatoms. 
It is clear from this that their structural analyses have no consistency by 
themselves because they added a dimer only to Fig.\ 1 and not to Fig.\ 2. 
Also, they added the rest atom to the position where its backbonds must suffer 
from the strong distortion to increase a stress energy. 
Although it is necessary to carry out STM observations with negative 
sample bias voltage $V_{s}$ to identify the locations of rest 
atoms~\cite{hamers86}, all of their STM images were obtained at 
$V_{s} = 2.0$ V only~\cite{gu95}. 
Based only on such STM images, nobody can discuss whether rest atoms are 
really present at surface sites which look dark in the empty-state images. 
For this reason, Fig.\ 1 of Ref.~\onlinecite{zhao98} is dubious. 

In addition, although they stated that Fig.\ 2 of
Ref.~\onlinecite{zhao98} was beyond our classification scheme, one can
find the corresponding shift in the location of the $(5,5)$ pattern in
Fig.\ 3 of Ref.~\onlinecite{itoh93} and, hence, their criticism is
simply incorrect. 
Actually, based on our assumptions and with the help of a mirror symmetry as 
we discussed in Ref.~\onlinecite{itoh93}, it is easy to deduce the
atomic structures of the $(n,n)$ patterns for any integer values of $n$
($0 \leq n \leq 6$), including the structure model they depicted in
Fig.\ 2; it merely confirmed the validity of our scheme and the
assumptions we adopted in Ref.~\onlinecite{itoh93}.

According to their explanation, Fig.\ 4 was depicted based on the STM image 
of Fig.\ 6(a) in Ref.~\onlinecite{gu95}, in which the two $3\times3$
structures are found apart from each other with their adjacent sites
being either covered by the contaminations or made of the vacancies. 
Only with such a STM image, and without any filled-state images, again 
it is impossible to deduce the atomic structure correctly.

Finally, I will point out that a surface structure appearing after annealing 
depends on the annealing schedule due to an entropic effect~\cite{itoh97}, 
so that their argument in deducing the energy difference between the 
$9\times9$ structure and the $5\times5$ structure is insufficient to 
derive any definite conclusions.

In summary, firstly since they did not state any criteria to fulfill 
systematic analyses, their structure models have no consistency among them. 
Secondly since they did not observe any filled-state STM images, their 
results have no reliability on the presence or the location of rest atoms. 
Thirdly, they criticized the failure of our classification scheme based on the 
observation of the structure which is easily deduced 
by our classification scheme and by our assumptions and, hence, their 
criticism is simply incorrect. 
And finally they overlooked the importance of the entropic effect which is 
involved in the determination of atomic structures in an annealing process.

\end{document}